\documentclass[11pt]{article}
\usepackage{psfig}

\def\lesssim{\mathrel{\mathpalette\vereq<}}
\def\gtrsim{\mathrel{\mathpalette\vereq>}}
\makeatletter
\def\vereq#1#2{
\lower3pt\vbox{\baselineskip1.5pt \lineskip1.5pt
\ialign{$\m@th#1\hfill##\hfil$\crcr#2\crcr\sim\crcr}}}
\makeatother

\begin{document}
\begin{titlepage}
\begin{center}
\hfill    CERN-TH/2001-187\\
\hfill    IFIC/01-36\\
~{} \hfill hep-ph/0107186\\

\vskip 1cm

{\large \bf Solving the Solar Neutrino Puzzle with KamLAND and Solar Data}

\vskip 1cm

Andr\' e de Gouv\^ ea$^{1}$ 
and Carlos Pe\~na-Garay$^{2}$

\vskip 0.5cm

{\em $^{1}$ Theory Division, CERN, CH-1211 Geneva 23, Switzerland.}

\vskip 0.3cm

{\em $^{2}$ Instituto de F\'{\i}sica Corpuscular, Universitat de
Val\`encia - C.S.I.C. \\ Edificio Instituto de Paterna, Apt. 22085, 46071
Val\`encia, Espa\~na}
\end{center}

\vskip 3cm

\begin{abstract}
We study what will be learnt about the solar neutrino puzzle
and solar neutrino oscillations once the data from the KamLAND reactor neutrino 
experiment (soon to become available) are combined with those from the current solar 
neutrino experiments.
We find that, in agreement with previous estimates,
if the solution to the solar neutrino puzzle falls on the LMA region,
KamLAND should be able to ``pin-point'' the right solution 
with unprecedented accuracy after a few years of data taking. Furthermore,
the light side ($\theta<\pi/4$) can be separated from the dark side ($\theta>\pi/4$)
at the 95\% confidence level (CL) for 
most of the LMA region allowed by the current data at the 99\%~CL, 
while the
addition of the 
KamLAND data need not improve our ability to limit a sterile component in ``solar'' 
oscillations. If KamLAND does not see an oscillation 
signal, the solar data would point to 
the LOW/VAC regions, while the SMA region would still lurk at the two sigma~CL, 
meaning we would probably have to wait for Borexino data in 
order to finally piece the solar neutrino puzzle. 
\end{abstract}

\end{titlepage}

\newpage
\setcounter{footnote}{0}
\setcounter{equation}{0}
\section{Introduction}

The 30+ year old solar neutrino puzzle constitutes, together with the atmospheric
neutrino puzzle, the only clean evidence for physics beyond the Standard Model.
What started off as a discrepancy between the expected and
experimentally measured solar 
neutrino flux at the Chlorine experiment in the Homestake mine
(\cite{Cl}, and references therein), 
has now turned into, when the state-of-the-art SuperKamiokande \cite{SuperK}
and SNO experiments \cite{SNO} are
combined, strong evidence that active neutrinos other than $\nu_e$ 
({\it i.e.,}\/ $\nu_{\mu}$ and/or $\nu_{\tau}$) are part of the solar neutrino flux
\cite{SNO}.

However, when the solar neutrino puzzle is taken as evidence for neutrino oscillations,
many issues remain. First, the solar neutrino data is yet to
provide unambiguous evidence for oscillations, and even the evidence for neutrino 
conversion requires the combination of two different experiments (this issue will
soon be resolved after SNO measures neutral current events). Second, the current
data is still unable to establish what is the generic value of the neutrino oscillation
parameters \cite{Lisi_SNO,BGG_SNO,BCGK_SNO,CSS_SNO}. 
Even if the recently disfavoured small mixing angle (SMA) region is
discarded, there are still a few disjoined values of $\Delta m^2$ which 
properly fit all the data. The (preferred) large mixing angle (LMA) region
contains values of $10^{-5}~{\rm eV^2}\lesssim\Delta m^2\lesssim 10^{-3}$~eV$^2$
and $0.2\lesssim\tan^2\theta\lesssim 3$, while the ``low
probability'' (LOW), quasi-vacuum (QVO), and just-so (VAC) regions contain
$10^{-11}~{\rm eV^2}\lesssim\Delta m^2\lesssim 10^{-6}$~eV$^2$ and
$0.1\lesssim\tan^2\theta\lesssim 10$. Finally, it is still not clear whether
the $\nu_e$ is oscillating into an active neutrino or a sterile neutrino. While the
current data strongly disfavours pure $\nu_e\leftrightarrow\nu_{s}$ oscillations 
($\nu_s$ is a $SU(2)\times U(1)_{Y}$ singlet, sterile neutrino), the possibility that 
$\nu_{e}$ oscillates into a strongly mixed active $\oplus$ sterile 
state is quite robust \cite{Barger_SNO,4neuts}.

The KamLAND reactor neutrino experiment \cite{KamLAND}, 
which is to start taking data soon, is 
sensitive to the LMA region of the solar neutrino parameter space. After a 
few years of data taking, it is capable of either excluding the entire region or
not only establishing 
$\nu_{e}\leftrightarrow \nu_{\rm other}$ oscillations, but also of measuring the 
oscillation parameters $(\tan^2\theta,\Delta m^2)$ with unprecedented
precision \cite{Barger_KAM,BS_KAM,MP_KAM,Kam_KAM}.  

Even in the case of a positive signal for oscillations at KamLAND, a few outstanding
issues will remain. Because KamLAND is 
sensitive to $\bar{\nu}_e$ charged current events only, it is unable to tell 
whether the $\nu_e$ oscillates into an active or a sterile state. Furthermore, 
because it is a ``short'' baseline experiment, it is insensitive to matter effects, and 
therefore cannot separate the ``light'' from the ``dark'' side of the parameter space 
\cite{dark_side} ({\it i.e.,}\/ tell $\theta$ from $\pi/2-\theta$). 

In this paper, we try to determine what extra information will be gained when the 
KamLAND results are combined with the current solar neutrino data collected by
Homestake, Sage, Gallex, GNO, Kamiokande/SuperKamiokande, and SNO 
\cite{Cl,SuperK,SNO,other_exp}. In particular,
we find that, for three years of KamLAND ``data:'' 
{\it (i)}~for a significant portion of the LMA region, we should be able to 
separate the light from the dark side at more than 95\% confidence level (CL), given
that the KamLAND data is precise enough to rule out maximal mixing with the
same precision; 
{\it (ii)}~if KamLAND sees a signal, we can establish that the $\nu_s$ component 
of the ``other'' neutrino is smaller than $\lesssim 0.6$ at the 99\%~CL. 
This bound, however,
is sensitive to the values of $\Delta m^2$ and $\tan^2\theta$ which are
to be measured at KamLAND. If KamLAND does not see a signal,
this upper bound is $0.78$ at the 99\%~CL.
{\it (iii)}~If KamLAND does not see a signal for oscillations, the solar data
will point to the LOW, QVO, or VAC regions, while the SMA region should still
be allowed at the $(2\div 3)$~sigma level.  

Our presentation is as follows. In the next section, we briefly review the KamLAND
experiment, and discuss the details of our simulations of KamLAND data and data 
analysis. In Sec.~3, we discuss the consequences of combining KamLAND and solar data 
if there is evidence for $\bar{\nu}_e$ disappearance at KamLAND, while in Sec.~4 we 
discuss the status of the solar neutrino puzzle in the advent that KamLAND sees no 
suppression of the reactor $\bar{\nu}_e$ flux. In Sec.~5 we summarise our results and 
conclude.

\setcounter{equation}{0}
\section{KamLAND experiment and ``data'' analysis}

The KamLAND detector, located inside a mine in Japan (in the site of the old Kamiokande
experiment), consists of roughly 1 kton of liquid scintillator surrounded by 
photomultiplier tubes. It is sensitive to the $\bar{\nu}_e$ flux from some 10+ reactors 
which are located ``nearby.'' The distances from the different reactors to the 
experimental site vary from slightly more than 80~km to over 800~km, while the 
majority (roughly 80\%) of the neutrinos travel from 140~km to 215~km (see, 
{\it e.g.,}\/ \cite{KamLAND} for details). KamLAND ``sees'' the antineutrinos by
detecting the total energy deposited by recoil positrons, which are
produced via $\bar{\nu}_e+p\rightarrow e^{+}+n$. The total visible energy corresponds to 
$E_{e^+}+m_{e}$, where $E_{e^+}$ is the kinetic energy of the positron and
$m_e$ the electron mass. The positron energy, on the other hand, is related to the
incoming antineutrino energy 
$E_{e^+}=E_{\nu}+1.293$~MeV up to corrections related
to the recoil momentum of the daughter neutron (1.293~MeV is the 
neutron--proton mass difference). KamLAND is expected to measure the visible energy 
with a resolution which is expected to be better than
$\sigma(E)/E=10\%/\sqrt{E}$, for $E$ in MeV \cite{KamLAND,Kam_KAM}. 

In order to simulate events at KamLAND, we need to compute the expected energy 
spectrum for the incoming reactor antineutrinos for different values of the neutrino
mass-squared differences and mixing angles.
 
The antineutrino spectrum which is to be measured at KamLAND depends on the power 
output and fuel composition of each reactor (both change slightly as a function 
of time), and on the cross section for $\bar{\nu}_e+p\rightarrow e^{+}+n$. For the results
presented here, we assume a constant chemical composition for the fuel of all reactors
(explicitly,  53.8\% of $^{235}$U, 32.8\% of $^{239}$Pu, 7.8\% of $^{238}$U, and
5.6\% of $^{241}$Pu, see \cite{Barger_KAM,chem_comp}). Effects due to the time variation
of the fuel composition
have been studied in \cite{MP_KAM}, and are small (although they have to be taken into
account in the ``real'' data analysis, of course).

The shape of energy spectrum of the incoming neutrinos can be
derived from a phenomenological 
parametrisation, obtained in \cite{shape_spectrum},
\begin{equation}
\frac{{\rm d}N_{\bar{\nu}_e}}{{\rm d}E_{\nu}}\propto e^{a_0+a_1E_{\nu}+a_2E^2_{\nu}},  
\label{eq:spectrum}
\end{equation}
where the coefficients $a_i$ depend on the parent nucleus. The values of $a_i$ for the
different isotopes we used are tabulated in \cite{shape_spectrum,MP_KAM}. These 
expressions are very good approximations of the (measured) reactor flux for values of 
$E_{\nu}\gtrsim 2$~MeV. 

The cross section for $\bar{\nu}_e+p\rightarrow e^{+}+n$ has been computed including 
corrections related to the recoil momentum of the daughter neutron in \cite{cross_sec}. 
It should be noted that the energy spectrum of antineutrinos produced at nuclear 
reactors has been measured with good accuracy at previous reactor neutrino experiments 
(see \cite{KamLAND} for references). For this reason, we will assume that the 
expected (unoscillated) antineutrino energy spectrum is know precisely. 
Some of the effects of uncertainties in the incoming flux on the determination of 
oscillation parameters have been studied in \cite{MP_KAM}, and are supposedly small.   

Finally, we have to include the effects of neutrino oscillations. Here, we will constrain
ourselves to the simplest two-level system, {\it i.e.,}\/ we assume 
$\nu_{e}\leftrightarrow\nu_{\rm other}$ oscillations governed by only one mass-squared
difference $\Delta m^2$ (which is the ``solar'' $\Delta m^2$) and one mixing angle
(we choose $\tan^2\theta$ as the physical variable in order to separate the 
light side ($\theta<\pi/4$) from the dark side ($\theta>\pi/4$) of the parameter space 
\cite{dark_side,Lisi3}). The reason for this is
that, according to \cite{Barger_KAM}, 
the KamLAND experiment should not have enough sensitivity to distinguish a nonzero
$U_{e3}$ from zero, given the current bound provided by the Chooz experiment 
\cite{Chooz}. This being 
the case, the (energy dependent) electron antineutrino survival probability
at KamLAND is
\begin{equation}
P(\bar{\nu}_e\leftrightarrow\bar{\nu}_e) =
1 - \sum_i f_i \sin^22\theta\sin^2\left(\frac{1.27\Delta m^2L_i}{E_{\nu}}\right),
\end{equation} 
where $L_i$ is the distance of reactor $i$ to KamLAND in $km$, 
$E_{\nu}$ is in GeV and $\Delta m^2$
is in eV$^2$, while $f_i$ is the fraction of the total neutrino flux which comes
from reactor $i$ (see \cite{KamLAND}), which we assume is constant as a function of
time.\footnote{This need not be the case. Indeed, different reactors shut down at
different times of the year, implying that $f_i$ does vary 
as a function of time. This effect, of course, will be considered in the analysis of
real experimental data.}

With all these ingredients at hand, we simulate KamLAND data for different values
of $\tan^2\theta$ and $\Delta m^2$ in 12 visible energy 
bins of width $500$~keV, restricting ourselves to neutrino energies above 
2~MeV\footnote{This corresponds to a visible energy of roughly 1.22~MeV.} and
below 8~MeV. The width of the bin is chosen according to the expected energy resolution,
the low energy cutoff is included in order to respect the validity of 
Eq.~(\ref{eq:spectrum}), and the high energy cutoff is included in order to avoid 
bins with too few events, such that a Gaussian approximation for the statistical errors
can be safely used. Finally, we define one ``KamLAND-year'' as the amount of time
it takes KamLAND to see 800 events with visible energy above 1.22~MeV. This is roughly
what is expected after one year of running (assuming a fiducial volume of 1 kton), 
if all reactors run at (constant) $78\%$ of their maximal power output \cite{KamLAND}.    
       
Our simulated data sets are analysed via a standard $\chi^2$ function,
\begin{equation}
\chi^2(\tan^2\theta,\Delta m^2)=\sum_{j=1}^{N_{\rm bin}} 
\frac{\left(N_j - T_j(\tan^2\theta,\Delta m^2)
\right)^2}{\left(\sqrt{N_j}\right)^2} + N_{\rm d.o.f},
\label{eq:chi2}
\end{equation}
where $N_j$ is the number of simulated events in the $j$-th energy bin, and
$T_j(\tan^2\theta,\Delta m^2)$ is the theoretical prediction for this number of 
events as a function of the oscillation parameters. $N_{\rm bin}=12$ is the total
number of bins, while $N_{\rm d.o.f}$ is the number of degrees of freedom. This is done
in order to estimate the statistical capabilities of an {\sl average} 
experiment.\footnote{see the Appendix A of \cite{seasonal} for a detailed 
discussion of this procedure.} 
An alternative option would be to include random statistical fluctuations in the 
simulated data. In this case, the appropriate definition of $\chi^2$ would be 
Eq.~(\ref{eq:chi2}) minus $N_{\rm d.o.f}$. Note that we assume statistical errors
only, and do not include background induced events. This seems to be a 
reasonable assumption, given that KamLAND is capable of tagging the $\bar{\nu}_e$
by looking for a delayed $\gamma$ signal due to the absorption of the recoil
neutron. However, ``geophysical'' $\bar{\nu}_e$ may prove to be irreducible background
at the low energy bins ($E_{\rm visible}\lesssim 2.6$~MeV) \cite{Kam_KAM}. For this 
reason, we discuss in the Appendix results of a 
modified analysis of KamLAND ``data,'' where
the events contained in the lowest energy bins are discarded.    

\setcounter{equation}{0}
\section{KamLAND sees evidence for $\bar{\nu}$ disappearance}

\begin{figure}
\centerline{
\parbox{0.7\textwidth}{\psfig{file=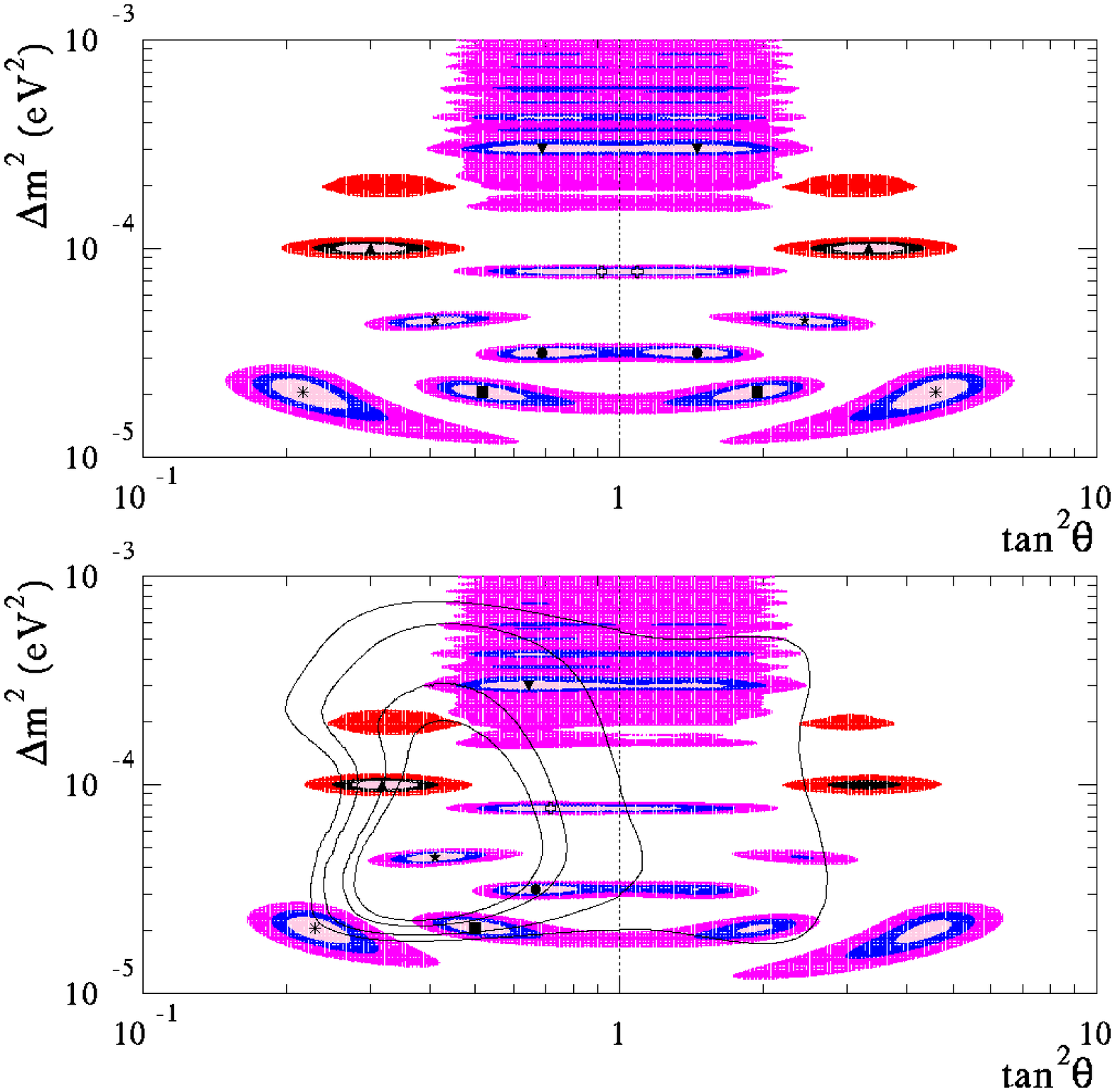,width=0.7\columnwidth}}
\parbox{0.3\textwidth}{\caption{TOP- Region of the 
$(\Delta m^2\times\tan^2\theta)$-parameter space allowed 
by three KamLAND-years of simulated data (see text) at the 90\%, three~sigma and 
five sigma confidence levels (CL), for different input values of $\Delta m^2$ and 
$\tan^2\theta$. The different symbols (star, square, circle, etc) 
indicate the best fit points.
BOTTOM- Same as TOP, except that the
current solar data is also included in the fit, assuming that the electro-type neutrino
oscillates into a pure active state. The line contours indicate the current
LMA region, defined for 2 d.o.f., at 90\%, 95\%, 99\%, and three sigma~CL.}
\label{fig:points}}}
\end{figure}

If the solution to the solar neutrino puzzle lies in the LMA region, KamLAND should
not only see a signal, but also determine $\tan^2\theta$, $\Delta m^2$ with
unprecedented precision \cite{Barger_KAM,BS_KAM,MP_KAM, Kam_KAM}. 
Fig.~\ref{fig:points}(top) depicts 
90\%, three sigma, and five sigma confidence level (CL) contours obtained for different
simulated input values of $\tan^2\theta$, $\Delta m^2$, after three KamLAND-years of 
``data.'' These contour are defined
by plotting contours of constant $\Delta\chi^2(\tan^2\theta,\Delta m^2)
\equiv\chi^2(\tan^2\theta,\Delta m^2)-\chi^2_{\rm min}$, while the CL are defined for
two degrees of freedom ($\Delta\chi^2=4.61, 11.83, 28.76$ for 90\%, 
three, and five sigma CL, respectively).

Three important comments are in order. 
First, as observed in \cite{BS_KAM}, for $\Delta m^2\gtrsim 10^{-4}$~eV$^2$, 
the determination
of $\Delta m^2$ is rather ambiguous. This is due to the fact that if 
$\Delta m^2$ is too large, the KamLAND energy resolution is not sufficiently high
in order to resolve the oscillation lengths associated with these values of
$\Delta m^2$ and $E_{\nu}$. 
Second, there is a discrete degeneracy in the
allowed parameter space, as $\theta$ cannot be separated from
$\pi/2-\theta$ ($\tan^2\theta$ from $\tan^{-2}\theta$). Third, although 
$\tan^2\theta$, $\Delta m^2$ can be very well determined, we have no information 
regarding whether the $\nu_e$ is oscillating into an active neutrino, a sterile
neutrino, or a linear combination of the two. The reason for this, as has been 
emphasised before, is that KamLAND sees only charged current events, and that
matter effects are completely negligible at KamLAND-like baselines.

Next, we add our simulated KamLAND data to the current solar data, including the
recent result published by SNO \cite{SNO}. The procedure for analysing the solar
data is the same as the one adopted in \cite{BGG_SNO}, where we refer the readers
for details. In particular, the analysis includes the event rates measured at
Homestake and SNO (``charged current'' data only), the average of the event
rates measured at Sage, Gallex and GNO, and the SuperKamiokande data, binned in 
energy for the day and night events. Furthermore, we include the theoretical errors on the
neutrino fluxes according to the results of BP2000 calculations \cite{BP2000}. 
We define a combined $\chi^2$ function in the straightforward way:
\begin{equation}
\chi^2_{\rm combined}(\tan^2\theta,\Delta m^2) = \chi^2_{\odot}(\tan^2\theta,\Delta m^2) 
+ \chi^2_{\rm KamLAND}(\tan^2\theta,\Delta m^2).
\label{eq:combined}
\end{equation}
Fig.~\ref{fig:points}(bottom) depicts 90\%, three sigma, and five sigma~CL 
contours obtained for different
simulated input values for $\tan^2\theta$, $\Delta m^2$, using the combined solar data
plus three KamLAND-years of ``data,'' assuming that the $\nu_e$ oscillates into a 
pure active state. As before, we define the confidence level contours by plotting 
contours of constant
$\Delta \chi^2$, this time computed with Eq.~(\ref{eq:combined}).  
Fig.~\ref{fig:points}(bottom) also depicts
the LMA region of the parameter which is allowed by current data at the 90\%, 
95\%, 99\%, and three sigma~CL.
\cite{BGG_SNO}.

A few important features should be readily noted. First, the size of the 90\% and
three sigma
contours {\sl in the light side} are almost identical when one compares 
Fig.~\ref{fig:points}(top) [KamLAND only] and Fig.~\ref{fig:points}(bottom) [KamLAND
plus solar], with the exception of the point located at a very large value
of $\Delta m^2=3\times10^{-4}$~eV$^2$ (indicated in Fig.~\ref{fig:points}(top) by an
up-side-down triangle). 
At this point, the number of separated ``islands'' allowed at the three sigma 
level decreases slightly.
The five sigma contours, on the other hand, are slightly reduced
when the solar data is included in the fit. This is particularly true when the
input values of the oscillation parameters are ``far'' from the current 
LMA best fit point (indicated by a star). 
Second, the ``mirror symmetry'' about the maximal mixing axis 
($\tan^2\theta=1$) is broken. Indeed, in most cases, the addition of the solar data
allows one to claim a ``strong hint,'' at more than 90\%~CL, that the $\nu_e$ is
predominantly light. At five sigma, however, none of the points depicted in
Fig.~\ref{fig:points} allows for a ``discovery'' of this (fundamental) fact.
One curious feature is that, for very small values of $\Delta m^2$, the five sigma~CL
contour may be smaller in the light side than in the dark side, while the best
fit point is well defined to be in the light side. This behaviour is most clear in the
point marked with an asterix, and is due to the fact that the region below 
$\Delta m^2\lesssim1\times 10^{-5}$~eV$^2$ and $\tan^2\theta\lesssim 1$
is strongly disfavoured by the absence of a day-night effect at SuperKamiokande. 

In Sec.~3.1, we analyse in more detail the ability of the current solar
data to separate the light from the dark side of the parameter space, once KamLAND
data becomes available (and if an oscillation signal is observed). In 
Sec.~3.2, we discuss what a combined solar and KamLAND data analysis
will have to say about a sterile component in $\nu_e\leftrightarrow\nu_{\rm other}$
oscillations.

\subsection{Separating the light from the dark side}

\begin{figure}
\centerline{
\parbox{0.7\textwidth}{\psfig{file=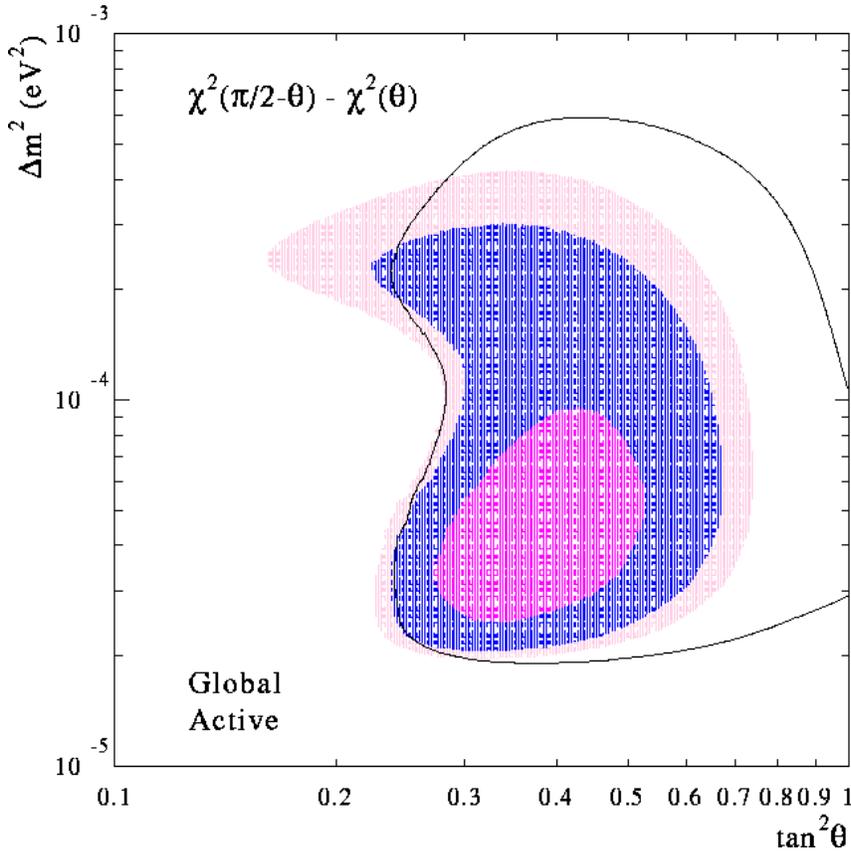,width=0.7\columnwidth}}
\parbox{0.3\textwidth}{\caption{Contours of constant $F\equiv|\chi^2_{\odot}(\theta)-
\chi^2_{\odot}(\pi/2-\theta)|$  
in the $(\Delta m^2\times\tan^2\theta)$-plane. The line contour corresponds 
to the LMA $99\%$ CL region, defined for 2 d.o.f. The regions correspond to 
$4.61<F<5.99$, $5.99<F<9.21$, $F>9.21$ from the outer-most to the inner-most contour.}
\label{fig:F}}}
\end{figure}

In order to address this issue, we compute the following ``discrimination function''
\begin{equation}
F(\tan^2\theta,\Delta m^2)\equiv|\chi^2_{\odot}(\tan^2\theta,\Delta m^2)-
\chi^2_{\odot}(\tan^2(\pi/2-\theta),\Delta m^2)|,
\end{equation}
which is defined for $0<\theta<\pi/4$. Fig.~\ref{fig:F} depicts contours of 
constant $F$ in the $(\tan^2\theta\times\Delta m^2)$-plane for 
$\Delta m^2>10^{-5}$~eV$^2$. $F$ is best interpreted in the
following way: if the best fit point obtained by analysing the solar data (combined or 
not with KamLAND data) is $(\tan^2\theta_*,\Delta m_*^2)$, than the mirror symmetric
point $(\tan^2(\pi/2-\theta_*),\Delta m_*^2)$ is ruled out at the $x$~CL, where the
value of $x$ is determined by $\Delta \chi^2=F(\tan^2\theta_*,\Delta m^2_*)$ for
two degrees of freedom. For example, if any of the points which lie on the $F=5.99$ 
contour in Fig.~\ref{fig:F} happens to be the best fit point, its symmetric
point on the ``other'' side will be ruled out at the 95\%~CL.   

Fig.~\ref{fig:F} also depicts the LMA $99\%$~CL region (defined for two d.o.f.). For 
a sizeable portion of it, 
$F$ is bigger than 5.99, meaning that we can expect to be able to separate
the dark from the light side of the parameter space at the 95\%~CL 
for a significant part of the currently preferred region of the solar 
neutrino parameter space. 
This conclusions depends, of course, on whether the KamLAND data by itself can 
discriminate the obtained best fit point from maximal mixing ($\tan^2\theta=1$). 
Note that, according to the current solar data, a five sigma ``discovery'' 
that the electron-type neutrino is predominantly light (or heavy), is not allowed by 
the current solar neutrino data. The largest value of $F=11.5$ occurs very close
to the best fit point indicated by the solar data. At this point, one can tell
the dark from the light side at the 99.68\%~CL.

\subsection{Constraining $\nu_{e}\leftrightarrow\nu_{\rm sterile}$ oscillations}

So far, we have assumed that $\nu_e$ oscillated into a pure active state 
$\nu_{a}$. Next we consider the more general case of $\nu_e\leftrightarrow
\nu_{\rm other}$ oscillations, where $\nu_{\rm other}$ is a linear combination 
of active and sterile neutrinos. Explicitly, we define the mixing angle $\zeta$
such that
\begin{equation}
|\nu_{\rm other}\rangle=\cos\zeta |\nu_{a}\rangle + \sin\zeta |\nu_{s}\rangle.
\label{zeta}
\end{equation} 
Note that we still assume that the $\nu_e$ is contained entirely in the two
mass eigenstates whose mass difference is responsible for solving the solar neutrino
puzzle.\footnote{In a four neutrino oscillation scheme, if we define $\nu_1$ and $\nu_2$
to be the mass eigenstates whose mass squared difference can be associated to 
$\Delta m^2_{\odot}$, this approximation corresponds to $|U_{e1}|^2+|U_{e2}|^2=1$,
while $\sin^2\zeta\equiv |U_{s1}|^2+|U_{s2}|^2$ \cite{4neuts}.} It is then 
possible to define oscillation probabilities,
compute event rates at the various experiments in terms of $\Delta m^2$, $\tan^2\theta$,
and $\sin^2\zeta$, and therefore compute 
$\chi^2_{\odot}(\tan^2\theta,\Delta m^2,\sin^2\zeta)$ \cite{4neuts}. 
Note that for KamLAND the
situation is trivial, as $\chi^2_{\rm KamLAND}$ is independent of $\sin^2\zeta$. 

In order to study what are the preferred values of $\sin^2\zeta$, we follow the
standard procedure of defining a one parameter $\chi^2$ function such that
\begin{equation}
\chi^2(\sin^2\zeta)\equiv \chi^2(\tan^2\theta_{\rm min},\Delta m^2_{\rm min},\sin^2\zeta),
\label{eq:chi2_zeta}
\end{equation}
where $\tan^2\theta_{\rm min},\Delta m^2_{\rm min}$ are the values of 
$\tan^2\theta,\Delta m^2$ which minimise 
$\chi^2$ for each value of $\sin^2\zeta$.

\begin{figure}
\centerline{
\parbox{0.7\textwidth}{\psfig{file=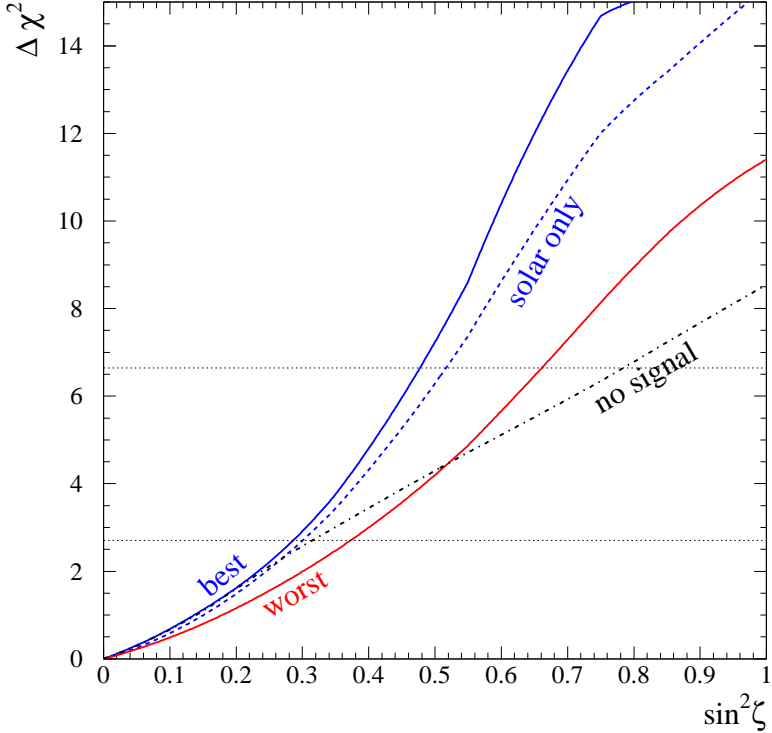,width=0.7\columnwidth}}
\parbox{0.3\textwidth}{\caption{$\Delta\chi^2(\sin^2\zeta)$ as a 
function of $\sin^2\zeta$, assuming that 
three KamLAND-years of ``data'' are consistent with two different point in the 
$(\Delta m^2\times\tan^2\theta)$-plane (solid lines, ``best'' and ``worst''), including 
only solar data in the fit (dashed line, ``solar only''), or assuming that
the three KamLAND-years of ``data'' do not contain any evidence for $\bar{\nu}_e$ 
oscillations (dot-dashed line, ``no signal''). The horizontal dotted lines indicate
the values of $\Delta\chi^2$ corresponding to a 90\% and 99\%~CL limits.}
\label{fig:sterile}}}
\end{figure}

Fig.~\ref{fig:sterile} depicts $\Delta\chi^2(\sin^2\zeta)\equiv\chi^2(\sin^2\zeta)-
\chi^2(\sin^2\zeta_{\rm min})$ as a function of $\sin^2\zeta$, for different scenarios.
The curve labelled ``best'' is obtained assuming that three KamLAND-years of
``data'' are consistent
with the current best fit point ($\tan^2\theta=0.41,~\Delta m^2=4.5\times10^{-5}$~eV$^2$,
indicated in Fig.~\ref{fig:points} by a star) to the
solar data assuming pure $\nu_e\leftrightarrow\nu_a$ oscillations, 
while the curve labelled ``worst'' is obtained assuming that 
three KamLAND-years of data are consistent with the best fit point in the would be LMA 
region assuming pure $\nu_e\leftrightarrow\nu_s$ 
($\tan^2\theta=0.69,~\Delta m^2=3.2\times10^{-5}$~eV$^2$,
indicated in Fig.~\ref{fig:points} by a circle). The dashed line labelled ``solar only'' 
is obtained if only the current solar data are included in the fit.

Three important features should be noted. First, $\sin^2\zeta=0$ is always the best fit
point. This is information contained in the current solar data by itself, of course.
Second, the solar data by itself sets an upper
bound $\sin^2\zeta< 0.30, 0.52$ at the $90\%, 99\%$CL, respectively. Third, if three 
KamLAND-years of data are consistent with the ``best'' point, the upper bounds 
on $\sin^2\zeta$ become slightly tighter 
($\sin^2\zeta< 0.28, 0.48$ at the $90\%, 99\%$CL), while if three 
KamLAND-years of data are consistent with the ``worst'' point the bounds actually
become looser ($\sin^2\zeta< 0.37,0.66$ at the $90\%, 99\%$CL). 

The reason for this behaviour is
the following: while KamLAND does not have the ability to discriminate active from
sterile oscillations, it affects $\chi^2(\sin^2\zeta)$ as defined in 
Eq.~(\ref{eq:chi2_zeta}) by modifying the values of $\tan^2\theta_{\rm min}$ and
$\Delta m^2_{\rm min}$. Roughly speaking, the KamLAND data almost ``fixes''
$\tan^2\theta_{\rm min}$ and $\Delta m^2_{\rm min}$ to the value chosen in the
simulated KamLAND data (see Fig.~\ref{fig:points}(top)), independent of $\sin^2\zeta$. 
At ``best,'' this phenomenon is quite visible: it forbids one to change the value of 
$\tan^2\theta$ and $\Delta m^2$ in order to obtain a better fit, and the upper bound
on $\sin^2\zeta$ becomes stronger. At ``worst,'' a different manifestation of the
same thing is present: the point ``worst'' is not particularly preferred by the
current solar data, but the presence of KamLAND ``forces'' it to be the best fit point.
It happens that the increase (with respect to the current best fit point)
in $\chi^2_{\odot}$ is larger at $\sin^2\zeta=0$ than at $\sin^2\zeta=1$, 
and for this reason the discrimination between pure active and pure sterile 
oscillations is poorer than the ``solar only'' case. 
%

\setcounter{equation}{0}
\section{KamLAND sees no evidence for $\bar{\nu}$ disappearance}

If the solution to the solar neutrino puzzle does not lie in the LMA region,
the KamLAND data should be consistent with the hypothesis that the electron-type
antineutrinos do not oscillate. Three KamLAND-years of ``data'' consistent with
no oscillations  
prove enough to rule out the entire LMA region, at more than three sigma~CL
(see \cite{KamLAND, Kam_KAM}).  

\begin{figure}
\centerline{
\parbox{0.7\textwidth}{\psfig{file=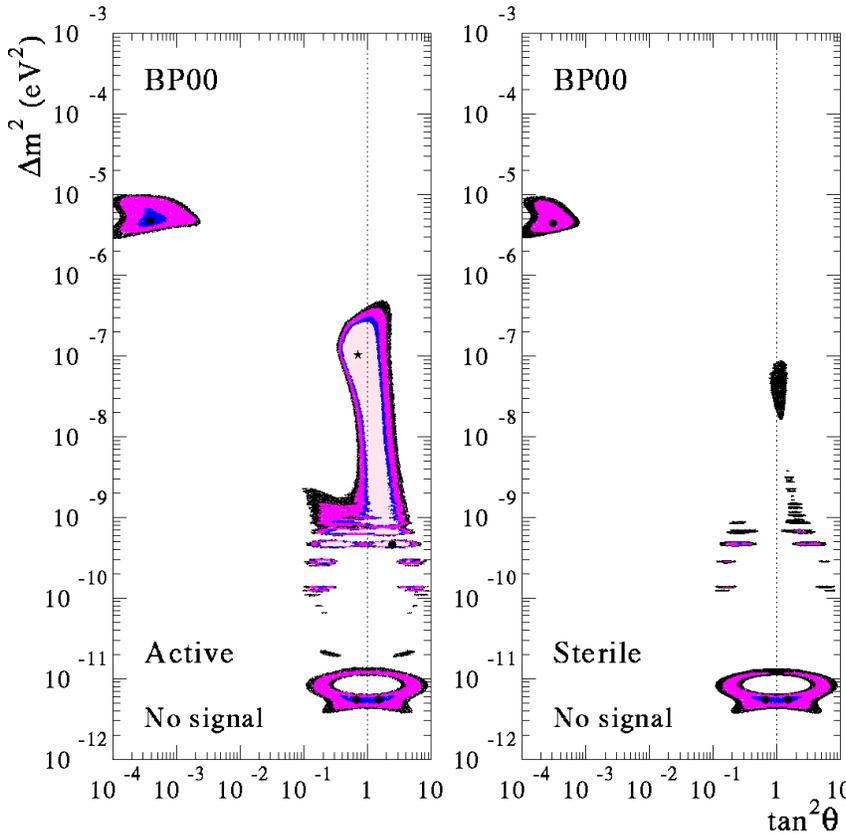,width=0.7\columnwidth}}
\parbox{0.3\textwidth}{\caption{Region of the 
$(\Delta m^2\times\tan^2\theta)$-parameter 
space allowed by the current solar data plus three KamLAND-years of 
simulated data at the $90\%$, 
$95\%$, $99\%$, and three sigma~CL (for three degrees of freedom), 
assuming that KamLAND sees no evidence for neutrino oscillations. The left panel
is for $\sin^2\zeta=0$ (pure active oscillations), while the left panel corresponds
to $\sin^2\zeta=1$. The theoretical errors for the BP2000 neutrino fluxes are included
in the analysis. See \cite{BGG_SNO} for details on the treatment of the solar data.}
\label{fig:no_signal}}}
\end{figure}

Fig.~\ref{fig:no_signal} depicts the ``left-over'' parameter space if this were
the case. Note that, here, we are performing fits with three free parameters, 
($\Delta m^2,\tan^2\theta,\sin^2\zeta$), and Fig.~\ref{fig:no_signal} depicts 
the ($\tan^2\theta\times\Delta m^2$)-plane for two fixed values of $\sin^2\zeta=0,1$ 
(left and right, respectively). Confidence level contours are therefore computed for
three degrees of freedom.

As expected, Fig.~\ref{fig:no_signal} resembles Fig.~1 in \cite{BGG_SNO}, with
the LMA solution ``chopped off.'' Some quantitative differences are
worthy of comment. For example, the best fit point is in the LOW region, while the 
entire range of values of $\Delta m^2$ which connect the LOW region to the VAC region
(the QVO region) is allowed at more than $90\%$ CL. Furthermore, the SMA region, which
is allowed by the current data only at the $99\%$ CL (according to the analysis 
performed in the spirit of \cite{BGG_SNO}), would be allowed at the $95\%$ CL defined
for three degrees of freedom.  

If this happens to be the result obtained by the KamLAND experiment, we will
probably have to wait for the upcoming Borexino experiment \cite{Borexino}
in order to learn more about solar neutrino oscillations. In particular, the 
search for anomalous seasonal variations at Borexino should either exclude
or establish the VAC and part of the QVO region, in which case oscillations
parameters may be measured with reasonably good precision \cite{seasonal}. On the other
hand, the search for a day-night asymmetry at Borexino should either exclude or 
establish the LOW and part of the QVO solution, in which case oscillations
parameters may also be measured with reasonably good precision \cite{day_night}.

Finally, Fig.~\ref{fig:sterile} depicts $\Delta\chi^2(\sin^2\zeta)$ as a function of
$\sin^2\zeta$ if the KamLAND data does not see a signal for oscillations 
(dot-dashed line, labelled ``no signal''). 
In this case, one would be able to set an upper bound
$\sin^2\zeta<0.32~(0.78)$ at the 90\%~(99\%)~CL, while at three sigma even values of 
$\sin^2\zeta=1$ would be allowed (as can be directly confirmed by looking at 
Fig.~\ref{fig:no_signal}). 

\setcounter{equation}{0}
\section{Summary and Conclusions}

We have considered the consequences of combining the current solar neutrino data
with future KamLAND reactor data. The main motivation for doing this is that while
KamLAND can explore the entire LMA region of the solar neutrino solution space 
and measure the oscillation parameters with great precision, we will need extra input,
which can only be provided by solar neutrino experiments (at least for the time
being), in order to address two important issues: first, we would like to decide whether
the electron-type neutrino is predominantly light or heavy, {\it i.e.,}\/ decide
whether the solar angle is larger or smaller than $\pi/4$ (in the convention where
the mass-squared hierarchy of the mass eigenstates is fixed \cite{dark_side}). Second,
we would like to know whether the electron-type neutrino mixes with the other known, 
active states (the muon- and tau-type neutrino), or whether it is (also)
mixed with an unknown sterile neutrino state. It is worthwhile to recall that if
the LSND anomaly \cite{LSND} is also interpreted in terms of neutrino oscillations,  
sterile neutrinos {\sl exist}.

We find that the current solar data do not allow a ``discovery'' (at the five 
sigma level) signal for $\theta<\pi/4$ (or $>\pi/4$) if KamLAND sees a signal 
for $\bar{\nu}_e$ disappearance,  
while a two sigma strong evidence may be claimed for a large part of the
99\% CL LMA region, if one assumes pure active oscillations. The situation is
qualitatively similar if any ``amount'' of sterile mixing is present.\footnote{Curiously,
for small values of $\Delta m^2\sim 10^{-5}$~eV$^2$ and pure sterile oscillations, 
the data prefers values of $\theta$ in the dark side ($\tan^2\theta>1$). Of course,
such scenario is severely constrained by the current solar data 
(see Fig.~\ref{fig:sterile}).} 
The big caveat, of course, is that the KamLAND data should be able to rule out 
maximal mixing ($\tan^2\theta=1$) with comparable confidence.

We also find that while KamLAND has no power to discriminate active from 
sterile oscillations, the KamLAND data will modify the current upper bound 
on a sterile neutrino component in ``solar'' oscillations. Curiously enough,
depending on the result obtained by KamLAND, this bound may improve or deteriorate
when compared to an analysis performed with the current solar data only.   

If KamLAND does not see evidence for $\bar{\nu}_e$ disappearance, the LMA 
region of the parameter space will be completely ruled out. The current solar 
data would, in this case, point to the LOW/QVO/VAC regions, while the SMA 
region would still be allowed at the two sigma level. If this is the case, data
from the Borexino experiment will probably be required if we want to finally
piece the solar neutrino puzzle. 

We would like to emphasise that most of the results presented here and summarised 
in the previous paragraphs are obtained from the {\sl current solar data,}\/
and do not depend on the KamLAND ``data.'' In particular, the conclusions we
reach should apply whenever the KamLAND data is ``precise enough'' to either
completely rule out the entire LMA region at a sufficiently high confidence
level, or measure $\tan^2\theta$ and $\Delta m^2$ values in the LMA region with
significantly higher precision than the solar data. Until that is the case, the 
lack of convincing absence or presence of signal at KamLAND will either 
slowly ``drown'' the LMA solution (if it does not contain the correct solution to 
the solar neutrino puzzle) or create other potential degeneracies in the parameter space  
(one could picture a ``best solar'' versus ``best KamLAND'' degeneracy 
problem). Incidently, this type of ``unresolved''
scenario may indeed be upon us when the first results from KamLAND are released, 
perhaps before the Summer of 2002 \cite{Kam_KAM}! 

Of course, as more solar data accumulates, the situation regarding the issues
studied here may change. In particular, the SNO detector is still to measure the 
``neutral current'' event rate \cite{SNO}, while more information will be obtained 
once the number of observed ``charged current'' events increases, and SNO is able to 
study the variation of the data sample as a function of time (day-night effect) 
and energy.

\section*{Acknowledgements}

We thank Concha Gonzalez-Garcia for many useful discussions and for 
collaboration during most of this work, and Alessandro Strumia for useful
conversations. CP-G also thanks the CERN Theory Division, where most of the work presented here was done, for its hospitality. This work was also supported 
by the Spanish DGICYT under grants PB98-0693 and PB97-1261, by the 
Generalitat Valenciana under grant GV99-3-1-01, and by the TMR network 
grant ERBFMRXCT960090 of the European Union and ESF network 86.

\appendix

\setcounter{equation}{0}
\setcounter{footnote}{0}
\section{More Conservative Analysis of KamLAND Data}

The KamLAND data sample may be plagued by an irreducible background of 
low energy $\bar{\nu}_e$, produced in the surrounding rock \cite{Kam_KAM}.
The ultimate way of reducing this background is to introduce a visible energy
threshold, above which the expected number of background events is negligible. 
According to \cite{Kam_KAM}, this cut should be placed at around 2.6~MeV.

In order to estimate the effect of this threshold, we repeat the analysis performed
in Sec.~3 by only including the highest nine energy bins (corresponding to
neutrino energies above 3.5~MeV, or visible energies slightly above 2.7~MeV). We again
define one KamLAND year as the time it takes KamLAND to observe, in the
absence of oscillations, 800 events 
{\sl with visible energy above 1.22~MeV}, such that the number of expected events
above 2.7~MeV is about 540.  

\begin{figure}
\centerline{
\parbox{0.7\textwidth}{\psfig{file=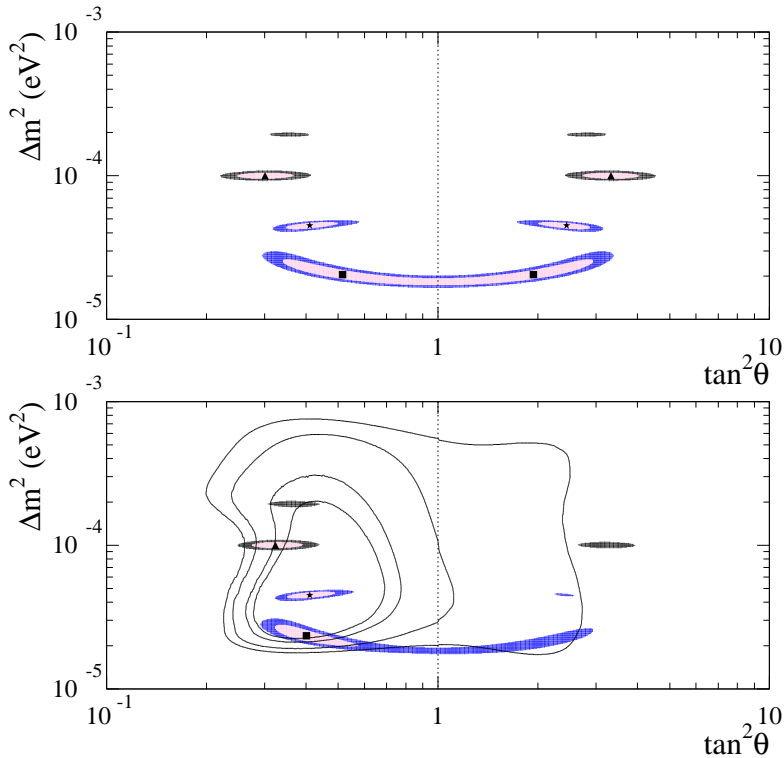,width=0.7\columnwidth}}
\parbox{0.3\textwidth}{\caption{Same as Fig.~\ref{fig:points}, 
when the three lowest energy bins at KamLAND are discarded from the ``data'' 
analysis. The confidence level contours correspond to 90\% 
and three sigma.}
\label{fig:points_new}}}
\end{figure}

Fig.~\ref{fig:points_new}(top)
depicts 90\% and three sigma CL contours obtained for different
simulated input values for $\tan^2\theta$, $\Delta m^2$, after three KamLAND-years of 
``data,'' excluding the three lowest energy bins. This is to be compared with
Fig.~\ref{fig:points}(top) in Sec.~3. Note that in Fig.~\ref{fig:points} we display
90\%, three, and five sigma confidence level contours. 

Fig.~\ref{fig:points_new}(bottom) depicts 90\% and three sigma~CL 
contours obtained for different
simulated input values for $\tan^2\theta$, $\Delta m^2$, using the combined solar data
plus three KamLAND-years of ``data'' minus the three lowest energy bins,
assuming that the $\nu_e$ oscillates into a pure active state.
This is to be compared with Fig.~\ref{fig:points}(bottom) in Sec.~3. 

Even if the CL contours are visibly bigger when one excludes the three lowest
energy bins (as expected), three ``KamLAND-years'' is more than enough to obtain 
a clean ``measurement'' of $\Delta m^2$ and $\tan^2\theta$, at least at the three
sigma level. Of course, the problem of multiple solutions at large mass-squared 
differences now plagues even smaller values of $\Delta m^2$, while the precision
with which one can determine oscillation parameters at $\Delta m^2$ close to
10$^{-5}$~eV$^2$ is significantly worse, as expected. Note that, 
as we advertise in Sec.~3.1,
the ability of the solar data to distinguish the light and dark sides does not
depend on how well KamLAND can measure the oscillation parameters. 
Furthermore, the effect of the solar data in the 
analysis is more pronounced in the light side, especially at $\Delta m^2$ close to
10$^{-5}$~eV$^2$. For example, while the KamLAND
result is unable to exclude maximal mixing at 90\%~CL at the
simulated point correspoding to $\Delta m^2=2\times 10^{-5}$~eV$^2$ (labeled by 
a square), the inclusion of the solar data significantly reduces the 90\%.

\end{document}